\documentclass[10pt,letter]{article}
\usepackage{opex3}
\usepackage{cite}

\begin{document}

\title{Opto-thermal analysis of a lightweighted mirror for solar telescope}

\author{Ravinder K. Banyal,$^*$  B. Ravindra, and S. Chatterjee}

\address{Indian Institute of Astrophysics, Bangalore, \\ India 560034}

\email{$^*$banyal@iiap.res.in} 



\begin{abstract}
 In this paper, an opto-thermal analysis of a moderately heated lightweighted solar telescope mirror is carried out using 3D finite element analysis (FEA). A physically realistic heat transfer model is developed to account for the radiative heating and energy exchange of the mirror with surroundings. The numerical simulations show the non-uniform temperature distribution and associated thermo-elastic distortions of the mirror blank clearly mimicking the underlying discrete geometry of the lightweighted substrate. The computed mechanical deformation data is analyzed with surface polynomials and the optical quality  of the mirror is evaluated with the help of a ray-tracing software. The thermal print-through distortions are further shown to contribute to optical figure changes and  mid-spatial frequency errors of the mirror surface. A comparative study presented for three commonly used substrate materials, namely, Zerodur, Pyrex and Silicon Carbide (SiC) is relevant to vast area of large optics requirements in ground and space applications.
\end{abstract}


\section{Introduction}
Large telescopes provide higher spatial resolution and improved sensitivity to probe exotic and unknown physical processes taking place in the Universe. Since last four hundred years, the telescope aperture size has gradually increased,  doubling almost in every 50 years period \cite{ren}. For night time astronomy, this impressive growth  will continue in the new millennia as the work on few notable 20-40 m class giant telescope projects is already in progress \cite{gil}. The evolution of solar telescope, on the other hand, has followed a different trajectory \cite{luh}. The immense difficulty in managing the thermal loads and internal seeing caused by radiative flux has been the major hurdles in building large solar telescopes. The technology has now advanced to a stage where building large aperture solar telescopes is well within the reach. Efforts from various groups are now underway to build next generation 2-4 m class solar telescopes. Some notable solar facilities that are currently in different phases of development include: a 4m Advanced Technology Solar Telescope (ATST) to be build by the National Solar Observatory, USA \cite{kei}. The 4m European Solar Telescope (EST), under considerations from European Association for Solar Telescopes \cite{zuc}. A 2m class National Large Solar Telescope (NLST) proposed by  the Indian Institute of Astrophysics, India \cite{has}.  These telescopes will be equipped with adaptive optics and other state-of-the-art scientific instruments to facilitate diffraction-limited observations of the small scale features ($\sim$50 km) of the solar atmosphere. Meanwhile, the 1.6 m New Solar Telescope (NST) at Big Bear Solar Observatory, USA and the 1.5 m Gregor telescope built by a consortia led Kiepenheuer-Institute, Germany, have started providing high quality ground based observations of the sun\cite{sch,cao}.

Evidently, one of the most challenging tasks is to carefully handle the excessive heat generated by large radiation flux incident on the primary mirror. Equally important is the choice of high quality, thermally stable and rigid material capable of maintaining the optical figure over a specified range of observing conditions. The thermal response of the solar mirror is influenced by radiative heating of the mirror caused by direct light absorption and the temperature imbalance with the surroundings. Mirror heating adversely affects the final image quality of the telescope in two ways, as is explained below.

First and foremost, it is hard for a massive mirror having large heat capacity and high thermal resistance to attain a temperature balance with the surrounding air. Under such conditions, the temperature difference between the mirror surface and the ambient air leads to refractive index fluctuations along the beam path. This `mirror seeing' produces wavefront aberrations responsible for the image blurring. To preserve the image quality of the telescope, the effect of mirror seeing has to be minimized. This means the temperature difference between the mirror surface and the ambient air should not exceed beyond $\pm2^{\circ}$~C \cite{emd}. More crucially, the temperature uniformity across the mirror surface  has to be maintained within $\pm0.5^{\circ}$~C. Several approaches which include, air conditioning \cite{miy}, fanning the optical surface \cite{low}, ventilating mirror interior, resistive heating \cite{boh,gre} have been proposed to regulate the mirror temperature for ensuring high quality astronomical observations. Laminar air flow across the mirror homogenizes the temperature by disintegrating the `thermal plumes' closer to the surface. These methods also speed up the thermalisation process by improving the convective heat transfer rate between the substrate and the surrounding air.

Secondly, a poorly conducting mirror is differentially heated to create temperature inhomogeneities within the substrate volume. Thermally induced  material expansion or contraction  can significantly alters the optical quality of the telescope mirror. This problem can be eliminated  by choosing mirror substrates made of  ultra low expansion (ULE) material. In fact, the demand for high quality astronomical optics  has led to the development of many attractive low expansion materials such as Zerodur (Schott), fused quartz, Cervit, ULE (Corning), Clearceram (Ohara) and Astro-Sitall etc. These ULE materials usually have high heat capacity and low thermal conductivity, implying  they can retain a significant amount of heat, leaving the mirror prone to undesirable seeing effects and surface buckling under high thermal loads.

The severity of mirror heating problem scales adversely with the aperture size. To reduce the overall weight and large thermal inertia, the most preferred choice is to use lightweighted mirror structures that are easy to mount and do not deform significantly under gravity. Besides, the reduced mass also facilitates efficient and faster cooling of the mirror in harsh thermal conditions. For space applications, reducing the weight of the mirror has greater premium on the overall cost reduction of the payload.

The lightweight geometry is created by removing pockets of material from the blank without significantly compromising the rigidity and stiffness of the mirror \cite{ahm}. The extreme lightweighting is usually achieved by mechanical milling and acid etching. The cool air, circulated through the rear pockets can remove the excessive heat from the mirror very efficiently. The pocketed cell geometry introduces structural inhomogeneities within the mirror blank. The top faceplate (reflecting surface) of the mirror gets the mechanical support from the grid of discrete ribs. The effect of underlying geometry of the cellular structure usually manifests in the form of rib print-through patterns a) during the mechanical polishing of the faceplate and b) periodic temperature residuals on the mirror faceplate operating in extreme thermal conditions. While polishing, the lapping tool experiences greater reaction force from the rib locations to wear out more materials thus creating periodic thickness variations on the faceplate. The wavefront errors related to polishing print-through are well studied and there are ways to mitigate them \cite{gra,tam,you}.

In this paper we have examined the impact of relatively less explored  but nonetheless important `thermal print-through' issue associated with the lightweighted structure operating in thermally dynamic environment. The thermo-optic analysis, as it is increasingly recognized, is a vital component of the overall system design and necessary to predict and verify the satisfactory performance of the mirror \cite{seg,din}. Here, we have used FEA to solve the time dependent 3-D heat transfer equation to predict the mirror deformation arising from the thermally induced stress and temperature inhomogeneities of a lightweighted structure. For realistic numerical simulations, the location dependent solar flux and ambient heating model is incorporated into FEA analysis. Thermal response of three types of mirror materials is studied at varying ambient temperature conditions that may exist at various observatory locations. Temperature induced surface deformations were analyzed using  Zernike polynomials.  For optical analysis, the Zernike data was imported into Zemax -a ray tracing software.  A comparative study is carried out for three well accepted substrate materials in astronomical optics, namely, Zerodur, Pyrex and SiC.

\section{Mathematical formulation}
In order to study the primary mirror heating and the associated thermal stress/strain fields affecting the surface figure, each physical process should be identified. In the following we give a brief mathematical formulation of the heat transfer problem along with appropriate prescription for boundary and initial conditions.

\subsection{Heat transfer} In a simplified form, the telescope mirror can be considered as a 3D circular disk whose front face is exposed to intense solar irradiance $I(\textrm{W/m}^{2})$. The thin metallic coating on the faceplate absorbs about 10\%--15\% of radiant flux which is converted into heat. The temperature distribution inside the mirror changes as the heat conducts from front surface to the rear side. The mirror also exchanges heat (via convection and radiation) with ambient air, which is at a temperature $T_{\textrm{a}}$. A heat transfer equation can be derived  based on the principle of conservation of energy, solution of which gives the temperature distribution $T(x,y,z,t)$. The 2nd order partial differential equation relevant to the current studies for the conductive heat transfer in a solid can be expressed as \cite{inc}

\begin{equation}\label{eq1}
   \frac{\partial}{\partial x} \left(K\frac{\partial T}{\partial x}\right) +\frac{\partial}{\partial y} \left(K\frac{\partial T}{\partial y}\right)+\frac{\partial}{\partial z} \left(K\frac{\partial T}{\partial z}\right) =\frac{\partial T}{\partial t} +q(x,y,z;t)
\end{equation}
where $T$ is the temperature, $K=k/(\rho C_{p})$ is the thermal diffusivity of the system,  $C_{p}$ is the heat capacity, $k$ is the thermal conductivity, $\rho$ is the density and $q=Q/(\rho C_{p})$, where $Q(x,y,z;t)$ represents the source term for heat. In the present case there is no internal heat source, i.e. $Q(x,y,z;t)=0$ everywhere and at all instants of time. The heat flux enter the system through the surfaces, a fact which must be accommodated in the boundary condition.

For solving the temperature $T(x,y,z,t)$ we must assign:
\begin{itemize}
  \item[--]  the initial condition
  \begin{equation}\label{eq2}
  T(x,y,z;t=0)=T_i(x,y,x)  \;\;\;\;\; \mbox{and}
  \end{equation}

  \item[--] the boundary condition that the net incident flux, which is balance between the incident and emitted fluxes, is transmitted into the substrate and accounts for the temperature gradients on the domain surface $\Omega$. This condition is described by,
\end{itemize}

\begin{equation}\label{eq3}
  k \left (\nabla T\cdot \textrm{\textbf{n}}\right)|_{\Omega}=\left[q_{0}-h\left(T-T_{\textrm{a}}\right)-\varepsilon_{\scriptsize\textrm{s}} \sigma_{\scriptsize\textrm{R}}\left(T^{4}-T_{\textrm{a}}^4\right)\right]|_{\Omega}
\end{equation}

where $\textrm{\textbf{n}}$ is surface normal vector at any point on $\Omega$. In Eq.~(3) the \emph{l.h.s.} denotes the heat inflow due to temperature gradient on the surface, $q_{0}$ is incident flux due to solar radiation, $h\left(T-T_{\textrm{a}}\right)$ describes the convective heat flow to surroundings, where $T_{\textrm{a}}(t)$ is ambient temperature and $h$ is the heat transfer coefficient. The term  $\varepsilon_{\scriptsize\textrm{s}} \sigma_{\scriptsize\textrm{R}}\left(T^{4}-T_{\textrm{a}}^4\right)$ describes the radiative exchange with surroundings, where $\varepsilon_{\scriptsize\textrm{s}}$ is the emissivity of the surface and $\sigma_{\scriptsize {\textrm{R}}}$ is Stefan--Boltzmann constant.

\subsection{Model prescription for initial temperature, heat flux and ambient temperature}
\subsubsection{Initial temperature $T_{i}(x,y,z)$}
Due to poor thermal conductivity and asymmetric heating, the initial temperature distribution $T_{\textrm{i}}(x,y,z)$ inside the mirror cannot be assumed to be uniform, i.e., independent of $(x,y,z)$. In order to establish a definitive temperature field at $t=0$, simulations were carried out by subjecting the mirror to several day-time heating and night-time cooling cycles \cite{ban}. The resulting temperature distribution was used as initial $T_{i}$ for all subsequent studies.

\subsubsection{Surface boundary conditions for inward heat flux $q_{0}$}
The boundary prescription for inward heat flux $q_{0}$ in Eq.~(3) depends on the solar irradiance $I$(W/m$^2$) that varies between the sun-rise and the sun-set as \cite{ban}
\begin{eqnarray}\label{eq4}
  q_{0} &=& \gamma\;I \;\;\;\;\textrm{and}\\
  I&=&I_{0} \left[\sin(\varphi)\sin(\delta)+\cos(\varphi)\cos(\delta)\cos(H)\right]
\end{eqnarray}
 where $\gamma$ is absorption coefficient (fraction of radiant energy that is absorbed and converted into heat) of the metallic film on the reflecting surface,  $I_{0}$(W/m$^2$) is irradiance amplitude when sun is at zenith, $\delta$~(rad) is declination angle of the sun, $\varphi$~(rad) is latitude of the place and $H=\left(\pi/12\right)t$ is solar hour angle and $t$ is the local time.
\begin{figure}[htb]
\centering\includegraphics[width=7cm]{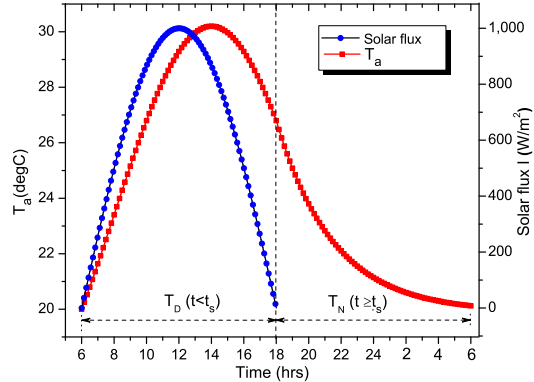}
\caption{Typical variations in ambient temperature (red curve, scale: vertical--left) and solar flux reaching the mirror surface. (blue curve, scale: vertical--right). The peaks of solar flux and ambient temperature are shifted by 2 hrs due to thermal time lag.}
\end{figure}

\subsubsection{A simple model for ambient temperature $T_{\textrm{a}}$}
In Eq.~(3), the heat loss or gain by free convection and radiation is driven by ambient temperature $T_{\scriptsize\textrm{a}}$. For convenience, we can split the ambient temperature into two parts, i.e., $T_{\textrm{a}}(t)=T_{\textrm{\scriptsize{D}}}(t)+T_{\textrm{\scriptsize N}}(t)$, where $T_{{\scriptsize \textrm{D}}}(t)$ and $T_{\textrm{\scriptsize N}}(t)$ describe the day-time ambient heating and night-time cooling, respectively. For cloud-free sky conditions, these two terms can be approximated using a simple model given by Gottsche and Olesen  \cite{got}

\begin{eqnarray}
  T_{{\scriptsize \textrm{D}}}(t) &=& T_{r}+T_{0}\cos\left[\frac{\pi}{\omega}(t-\tau_{m})\right] \;\;\; \textrm{for}\;\; t<t_{s} \\
  T_{{\scriptsize \textrm{N}}}(t) &=& T_{r}+T_{d} + \left \{ T_{0}\cos\left[\frac{\pi}{\omega}(t-\tau_{m})\right]-T_{d}\right \} \exp[-(t-t_s)/\kappa] \;\;\; \textrm{for}\;\; t\geq t_{s}
\end{eqnarray}
We use Eq.~(6) and Eq.~(7) as input to our FEA model to simulate a range of ambient temperature conditions typically existing at observatory locations.  One such realization for $10^{\circ}\textrm{C}\leq T_{\textrm{a}}(t)\leq 20^{\circ}\textrm{C}$ and solar flux $I$(W/m$^2$) is shown in Fig.~1. The assumed values for various parameters in Eq.~(6) and Eq.~(7) used to obtain a smoothly varying temperature curve in Fig.~1 are \cite{ban}: $T_{0}=6.8^{\circ}$C; $T_{r}=13.4^{\circ}$C; $T_{d}=-3.5^{\circ}$C; $\omega=12$hrs (the half period oscillation); $\tau_{m}=14:00$~hrs (the time when temperature maxima is reached); $t_{s}=18:00$~hrs (the time when temperature attenuation begins); $\kappa=3.5$~hrs (the attenuation constant). The ambient temperature $T_{\textrm{a}}$ shown in Fig.~1 continues to rise well past the solar noon. This is due to thermal time lag in terrestrial environment that exists between the peak of solar flux and the peak of ambient temperature.
\begin{figure}[htb]
\centering\includegraphics[width=10cm]{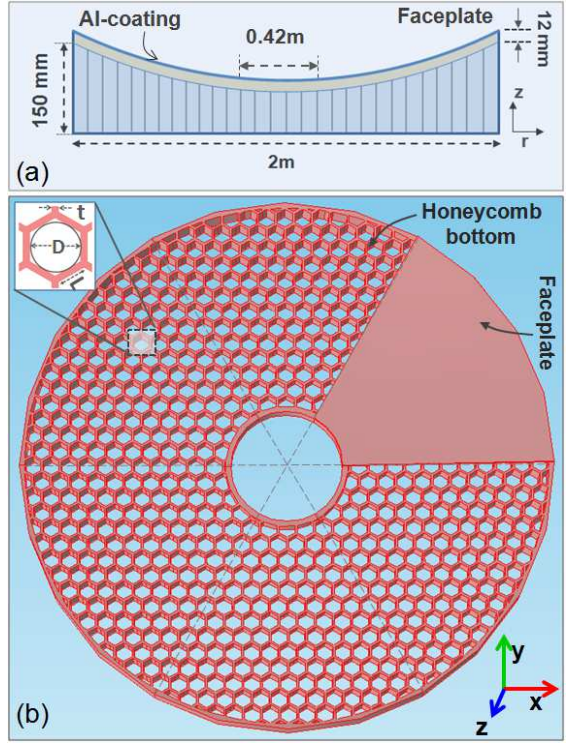}
\caption{Mirror geometry: (a) The side view and (b) the CAD model of the mirror with open back lightweighted honeycomb structure and 12 mm thick faceplate on top. Mirror has 6-fold symmetry about the z-axis. The faceplate is shown placed only on one of the 6 sectors of the mirror.}
\end{figure}
\subsection{Structural deformation}
The heat conducted to different parts of the mirror builds up thermal strain which results in structural deformation. For isotropic linear elastic solid, the thermal strain depends on coefficient of thermal expansion $\alpha$(K$^{-1}$), the instantaneous temperature $T(t)$ and the stress--free reference temperature $T_{\scriptsize\textrm{ref}}$ as:
\begin{equation}\label{eq7}
  \epsilon_{\scriptsize\textrm{th}}=\alpha(T-T_{\scriptsize\textrm{ref}})
\end{equation}
If $u$, $v$, and $w$ are the thermally induced deformation components in $x$, $y$ and $z$ direction, then the thermal strain can be completely specified in terms of $u$, $v$, and $w$ and their derivatives. For small displacements, the constitutive equations relating the components of normal and shear strain  to the deformation derivatives can be expressed as \cite{sad}:
 \begin{eqnarray}
   \epsilon_{xx} = \frac{\partial u}{\partial x}; \;\;  \epsilon_{xy} = \epsilon_{yx} = \frac{1}{2}\left(\frac{\partial u}{\partial y}+\frac{\partial v}{\partial x}\right)  \nonumber \\
  \epsilon_{yy} = \frac{\partial v}{\partial y}; \;\;  \epsilon_{yz} = \epsilon_{zy} = \frac{1}{2}\left(\frac{\partial v}{\partial z}+\frac{\partial w}{\partial y}\right) \\
   \epsilon_{zz} = \frac{\partial w}{\partial z} ; \;\;  \epsilon_{zx} = \epsilon_{xz} = \frac{1}{2}\left(\frac{\partial w}{\partial x}+\frac{\partial u}{\partial z}\right)\nonumber
 \end{eqnarray}
The displacements at the bottom boundary of the mirror were kept fully constrained, i.e., $u=v=w=0$, to avoid the rigid body motion.

\section{The mirror geometry}
Among several possibilities of lightweighted geometry, the open-back mirror configuration is highly suitable for solar telescopes \cite{sub}. It consists of a thin faceplate fused to an array of ribs formed by scooped-out material from the back. The pocket geometry can be circular,  triangular, hexagonal, square  or any other combination of these shapes. The resulting structure has side walls/ribs of certain thickness that supports the thin reflecting faceplate of the mirror from behind. Besides the mechanical support, the rib walls also provide the thermal flow path for heat to conduct away from the faceplate. The presence of discrete grid makes the heat removal rate to vary across the faceplate thus creating temperature impressions resembling the cellular geometry of the lightweighted structure.

Figure~2(a) shows the schematic of the parabolic mirror (conic constant =--1, focal length 4m) geometry chosen for the present studies. A similar design is also under consideration for the proposed NLST by India. The diameter of the mirror is 2m and the diameter of the inner hole is 0.42m. The honeycomb bottom comprising about 708 hexagonal cells is enclosed by 20~mm thick outer rim. The side length $L$ of the hex-cell (inset image in Fig.~2(b)) is 34.6~mm and the wall/rib thickness $t$ is 8~mm. The diameter  $D$ of the inscribed circle is 60~mm while the wall/rib depth varies from 148 mm for the outermost cells to 85~mm for the one at the center. The faceplate thickness is 12~mm. Compared to a solid geometry, the overall mass of the current lightweighted mirror is reduced by $\simeq68$\%. The main reason for choosing these hex-cell dimensions is because their structural stability has already been verified for the Gregor 1.5~m mirror telescope \cite{wes2}. The mirror has a 6-fold symmetry about the optical axis as seen in the CAD design shown in Fig.~2(b). The symmetry property is utilized to reduce the model size and computational requirements by selecting only one section of the mirror in the FEA studies.
\begin{figure}[htb]
\centering\includegraphics[width=6.8cm]{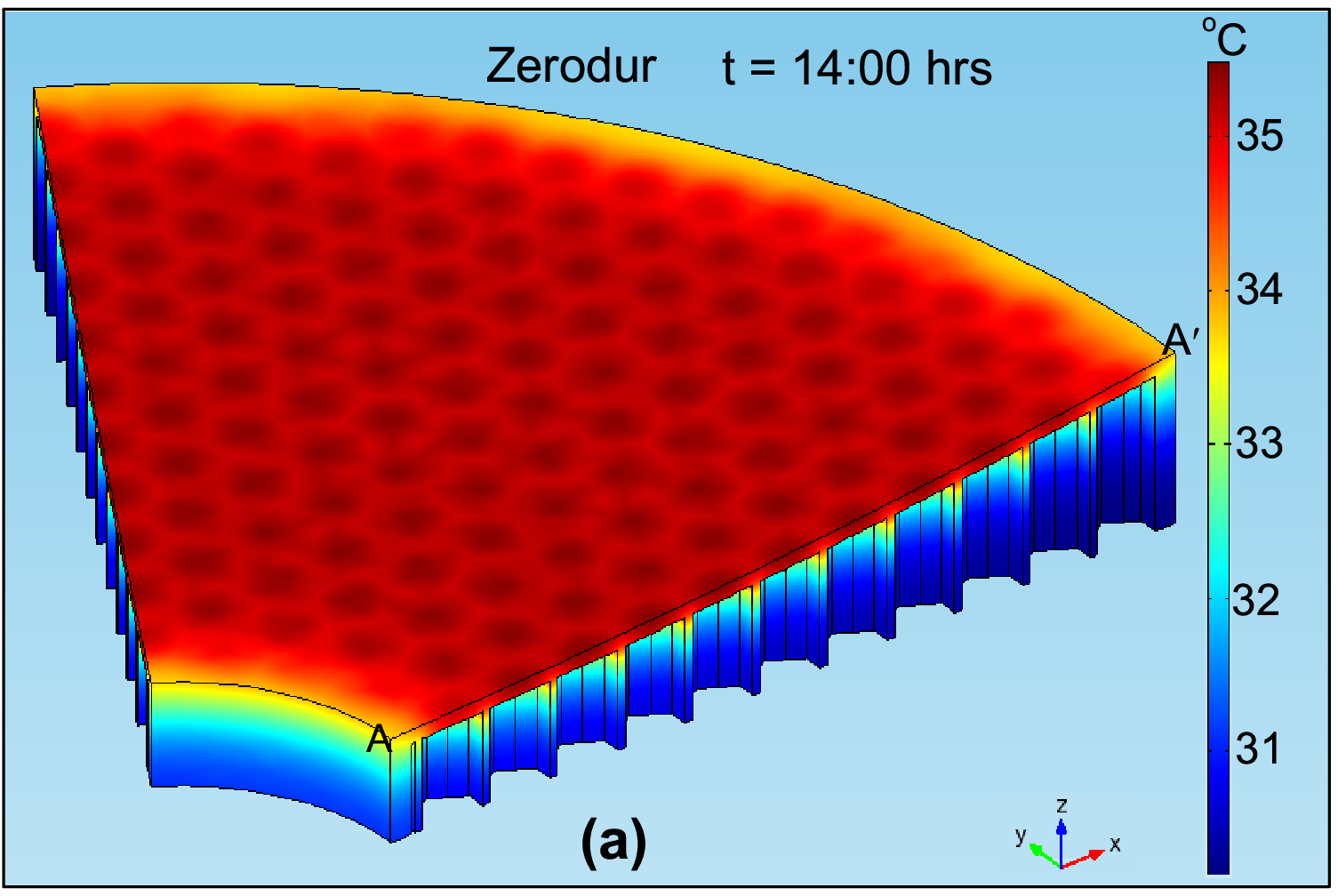}
\centering\includegraphics[width=6.2cm]{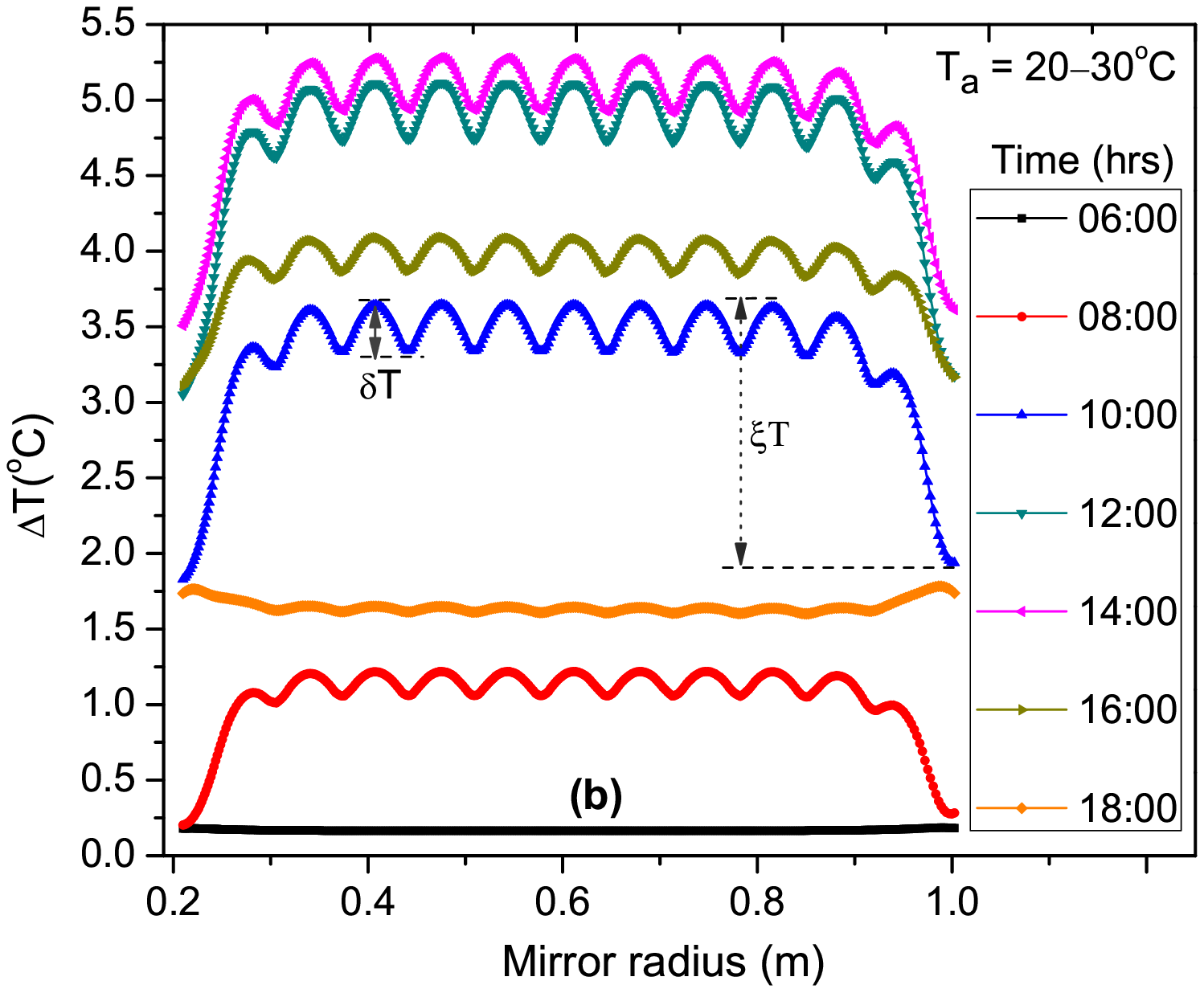}
\caption{The impact of daytime heating on the Zerodur mirror with ambient temperature varying  between 20--30$^\circ$C and $h=5$~W/m$^2$. (a) Temperature print-through pattern on the mirror faceplate subjected to daytime radiative heating. (b) Temperature variations along the radial line $AA^\prime$ of the mirror faceplate at different times.}
\end{figure}

\section{The FEA results}
The governing equations and associated boundary conditions for the heat transfer and the structural deformation problem are outlined in Section 2. The analytical solution for Eqs.~(1)--(3) and Eqs.~(8)--(9) exists only for simple and regular shapes.  In most practical cases where object geometry is highly complex, boundary conditions are complicated, material properties are temperature dependent and various physical parameters are coupled, problem then has to be solved by numerical methods. We used COMSOL Multiphysics as our FEA tool to solve the time-dependent 3D heat transfer and structural  problem. The segregated solver in COMSOL takes advantage of the direct coupling between heat transfer and structural model. That is, for  each time step ($\Delta t=100$ sec), the heat transfer module first solves for the temperature distribution of the mirror over 24 hours. Each temperature field is then passed onto the the structural analysis module to perform the stress-strain analysis. No data transformation was necessary since a common mesh is applicable to both thermal as well as structural models in COMSOL.

\begin{table}[htb]\small
 \centering\caption{Values of material properties used in FEA studies}\label{tab1}
 \begin{tabular}{l|l l l l |l}
         \hline
         Property                       &  \multicolumn{4}{c|}{Material}   & Units \\
                                        & Zerodur & SiC  & Pyrex  & Al     &\\ \hline
         CTE ($\alpha$)                 &-0.08    & 3.7  & 3.25   &  25    &  $10^{-6}\textrm{K}^{-1}$ \\
         Thermal conductivity ($k$)     & 1.46    & 118  & 1.14   & 237    &  $\textrm{WK}^{-1} \textrm{m}^{-1}$ \\
         Heat capacity ($C_{p}$)        & 821     & 672  & 750    & 910    &  $\textrm{Jkg}^{-1}\textrm{K}^{-1}$ \\
         Density ($\rho$)               & 2.53    & 3.20 & 2.23   & 2.70   & $10^{3}\cdot$kgm$^{-3}$ \\
         Emissivity ($\epsilon_{s}$)    & 0.9     & 0.9  & 0.95   & 0.05   &--\\
         Young's modulus ($E$)          & 90.3    & 414  & 64     &  70    & $10^{9}$ Pa \\
         Poisson's ratio ($\nu$)        & 0.24    & 0.14 & 0.2    & 0.33   & --\\ \hline
  \end{tabular}
\end{table}

A 3D CAD model of the mirror conforming to optical prescriptions is shown in Fig.~2. The reflecting surface of the faceplate was modelled using a 100~nm thick aluminium coating on the faceplate. The FEA model was run on the $1/6$th axisymmetric section of the mirror that was partitioned into 100644 tetrahedral mesh elements. The top surface of the mirror was continuously heated by time varying irradiated solar flux plotted in Fig.~1. The thermal and structural response was evaluated  at three different ranges of ambient temperature $T_{\textrm{a}}$ i.e., $0^{\circ}$C$-10^{\circ}$C, $10^{\circ}$C$-20^{\circ}$C and $20^{\circ}$C$-30^{\circ}$C. The front, back and the outer rim of the mirror were subjected to convective cooling with the heat transfer coefficient $h=5$~W/m$^2$ representing natural cooling and $h=15$~W/m$^2$ representing a moderately forced cooling. In addition, surface-to-ambient radiation condition was specified for the faceplate.  Also the symmetry boundary condition i.e., $k\left (\nabla T\cdot \textrm{\textbf{n}}\right)=0$, was imposed on the side edges of the mirror. In the FEA model, the mirror heating was \emph{`on'} (Eq.~(4)) between sunrise (6:00 hrs) and sunset (18:00 hrs) and \emph{`off'} for rest of the time. The standard values of the material parameters used in FEA studies are listed in Table~1 \cite{ban}.

\subsection{Thermal print-through and temperature nonuniformities}
 Figure~3(a) shows the temperature distribution within the mirror volume at 14:00 hrs when ambient temperature also reaches its peak. Apart from strong thermal gradients along axial direction, the appearance of  periodic temperature print-through pattern on the faceplate is clearly evident. The temperature excess $\Delta T=T-T_{\textrm{a}}$ between the mirror and the ambient air  is plotted in Fig.~3(b). The temperature nonuniformities on the mirror surface are quantified by two key variations. First, the cyclic undulations $\delta T$ (the temperature print-through effect) along the radial line $AA^\prime$ which is caused by differential heat flow from the faceplate to the underlying hex-cell structure.  Second, the trailing temperature $\xi T$ (edge-effect) which denotes the overall temperature difference between the mirror surface and the edges. This difference arises due to more efficient cooling from the walls of inner hole and the outer periphery of the mirror. According to data shown  for Zerodur in Fig.~3(b), the  $\xi T$ could be as high as 2$^\circ$C at noon. From the temperature distribution alone, we can loosely  identify $\xi T$ being associated with low order surface aberrations while $\delta T$  contributes to mid-spatial errors as explained in the next section.
\begin{figure}[htb]
\centering\includegraphics[width=7cm]{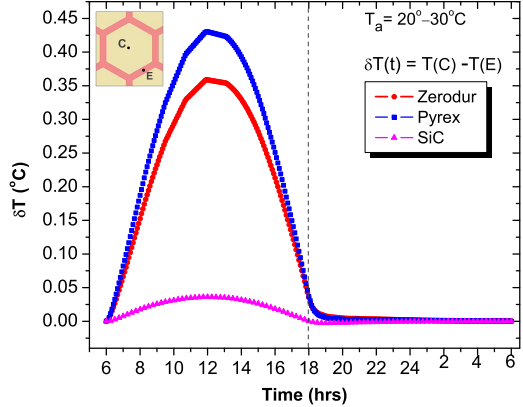}
\caption{Temperature difference between two reference points located at hexagonal cell-center (C) and cell-edge (E).}
\end{figure}
\begin{table}[htb]
 \centering\caption{The temperature difference for different glass material computed for heat transfer coefficient $h_{1}=5$~W/m$^2$ and $h_{2}=15$~W/m$^2$.}\label{tab1}\small
 \begin{tabular}{|l|c|c|c|c|c|c|}
         \hline
           &  \multicolumn{6}{c|}{$T_{\textrm{a}}$ range:  $20^\circ$C$-30^\circ$C}  \\
           \cline{2-7}
            Material & \multicolumn{2}{c|}{$\Delta T_{{\scriptsize\textrm{max}}}$}  & \multicolumn{2}{c|}{$\delta T_{{\scriptsize\textrm{max}}}$}   & \multicolumn{2}{c|}{$\xi T_{{\scriptsize\textrm{max}}}$} \\
                   & $h_1$ & $h_2$  & $h_1$ & $h_2$ & $h_1$ & $h_2$ \\  \hline
         Zerodur   & 5.38  & 2.56   & 0.36& 0.20   & 2.0  &  0.9  \\
         Pyrex     & 6.07  & 2.82   & 0.43& 0.21   & 2.0  &  1.0  \\
         SiC       & 2.56  & 1.02   & 0.04& 0.02   & 0.5  &  0.25   \\ \hline
        \end{tabular}

\end{table}

\begin{figure}[htb]
\centering\includegraphics[width=6.2cm]{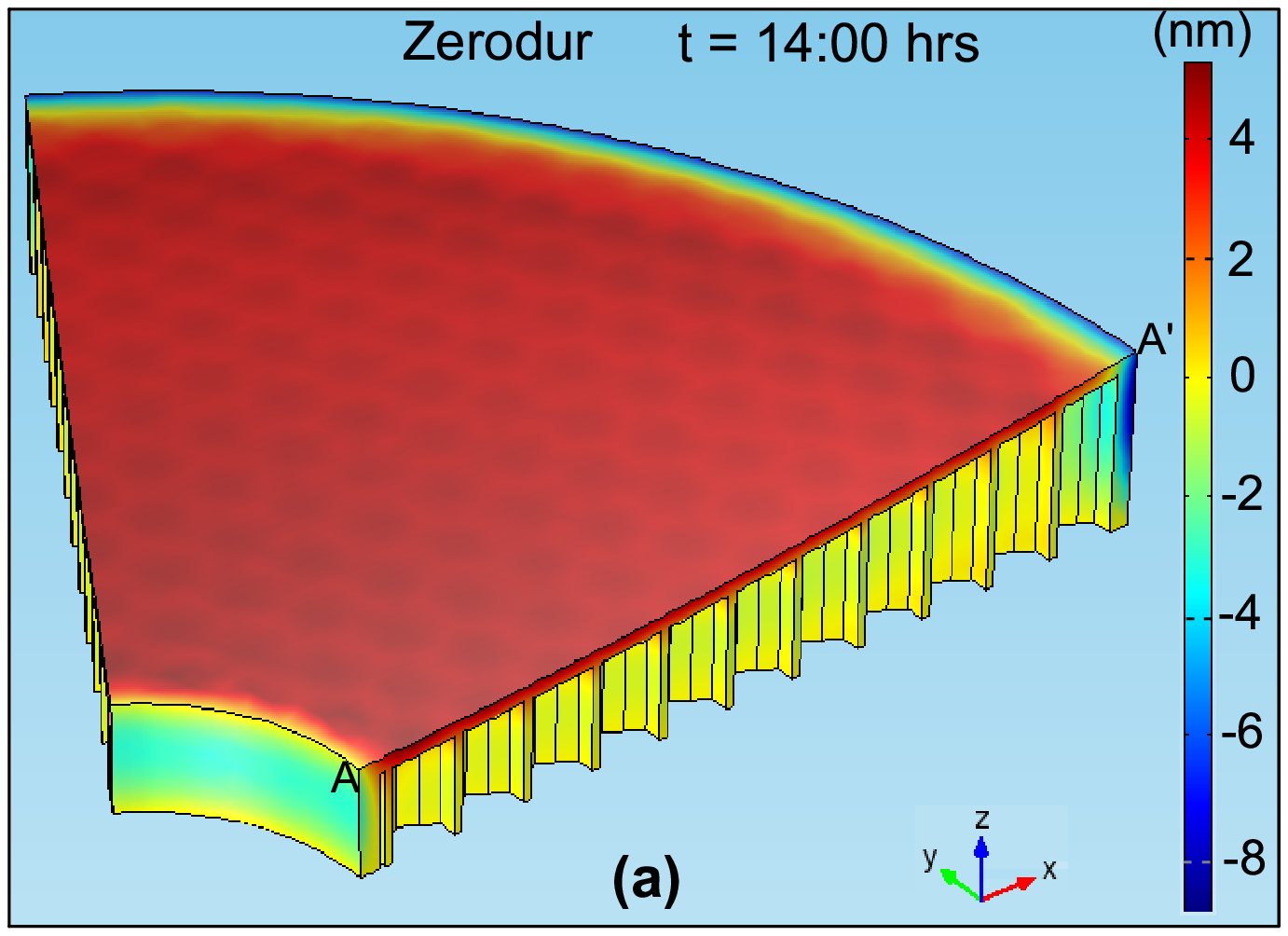}
\centering\includegraphics[width=5.6cm]{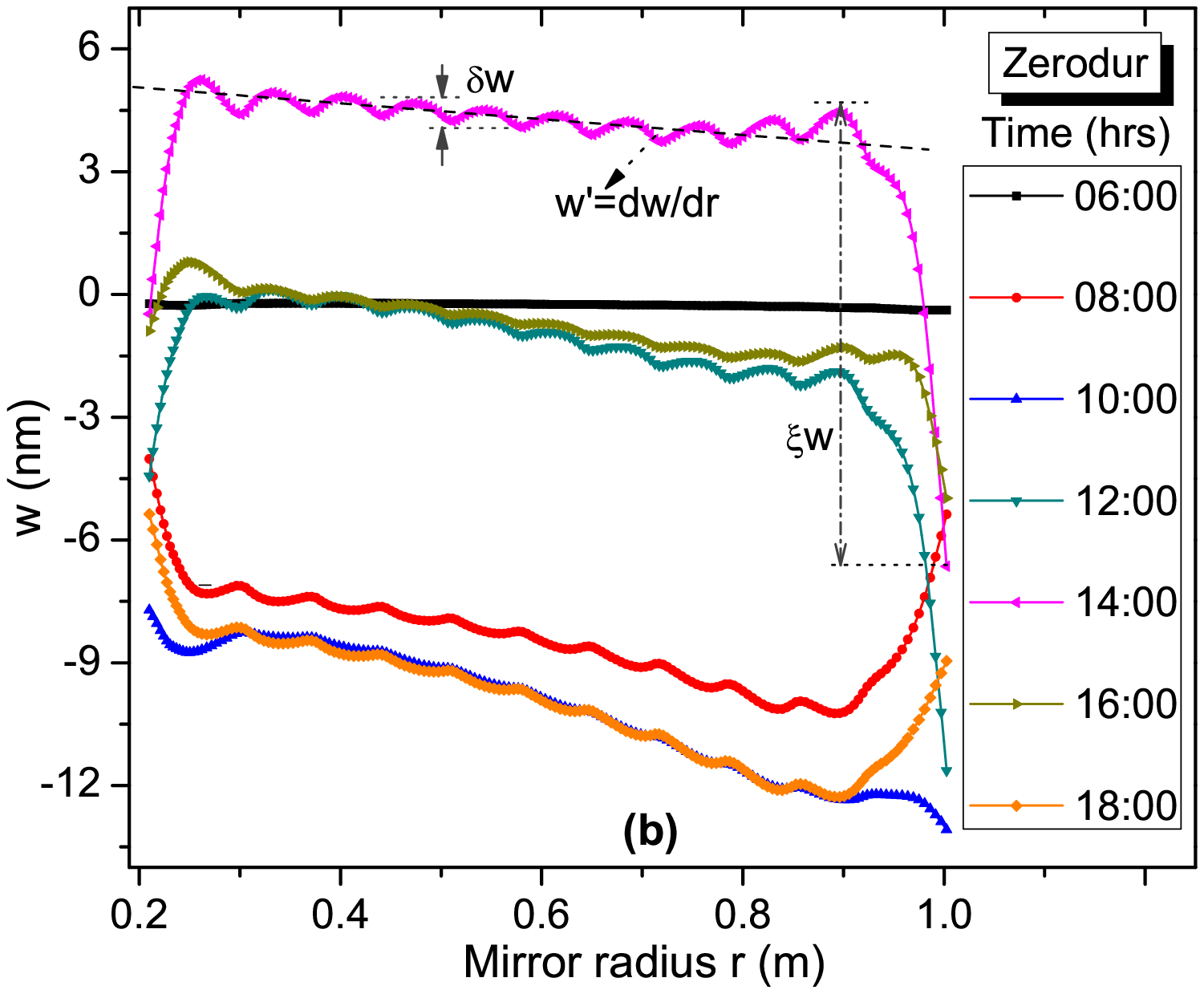}\\
\centering\includegraphics[width=6.0cm]{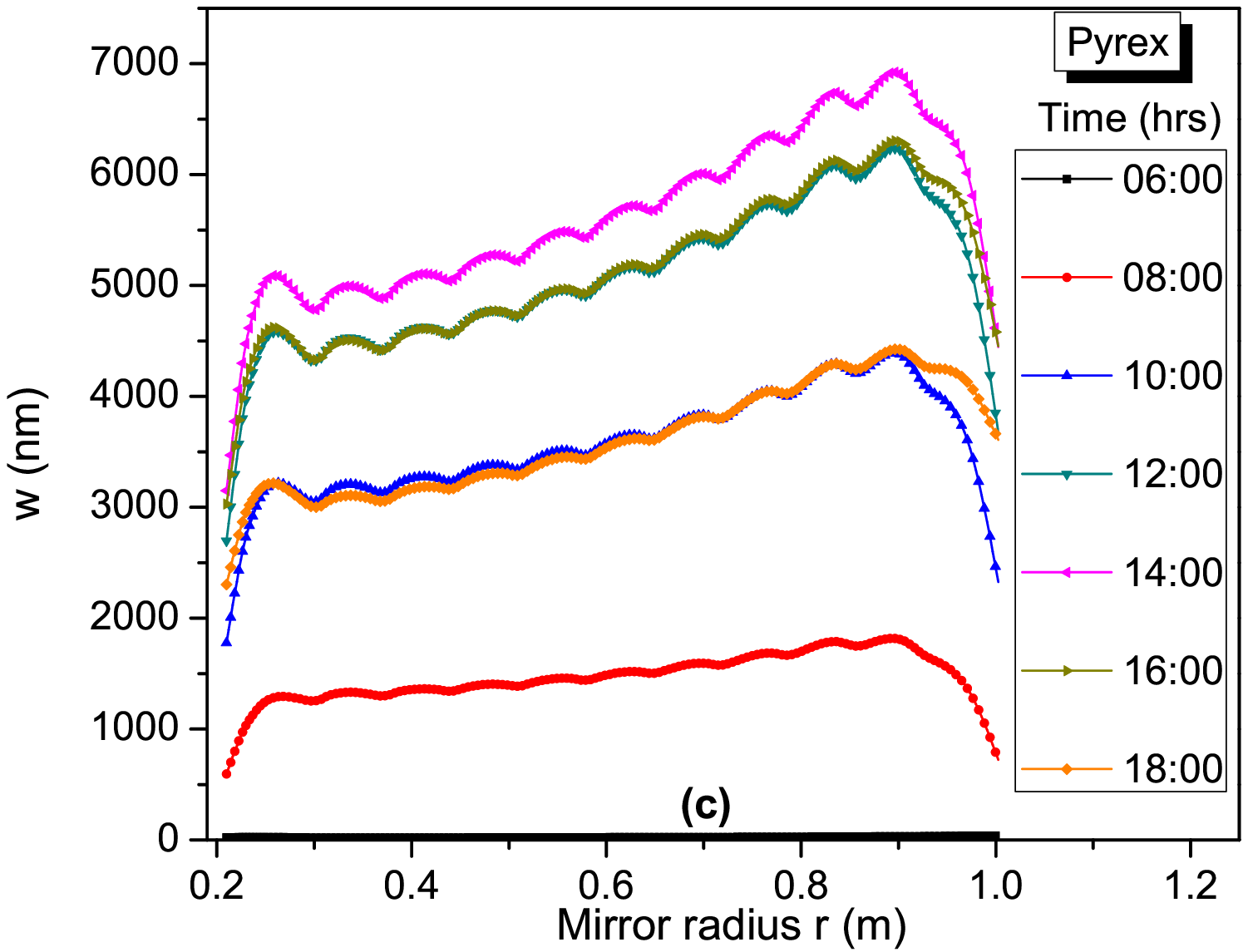}
\centering\includegraphics[width=6.0cm]{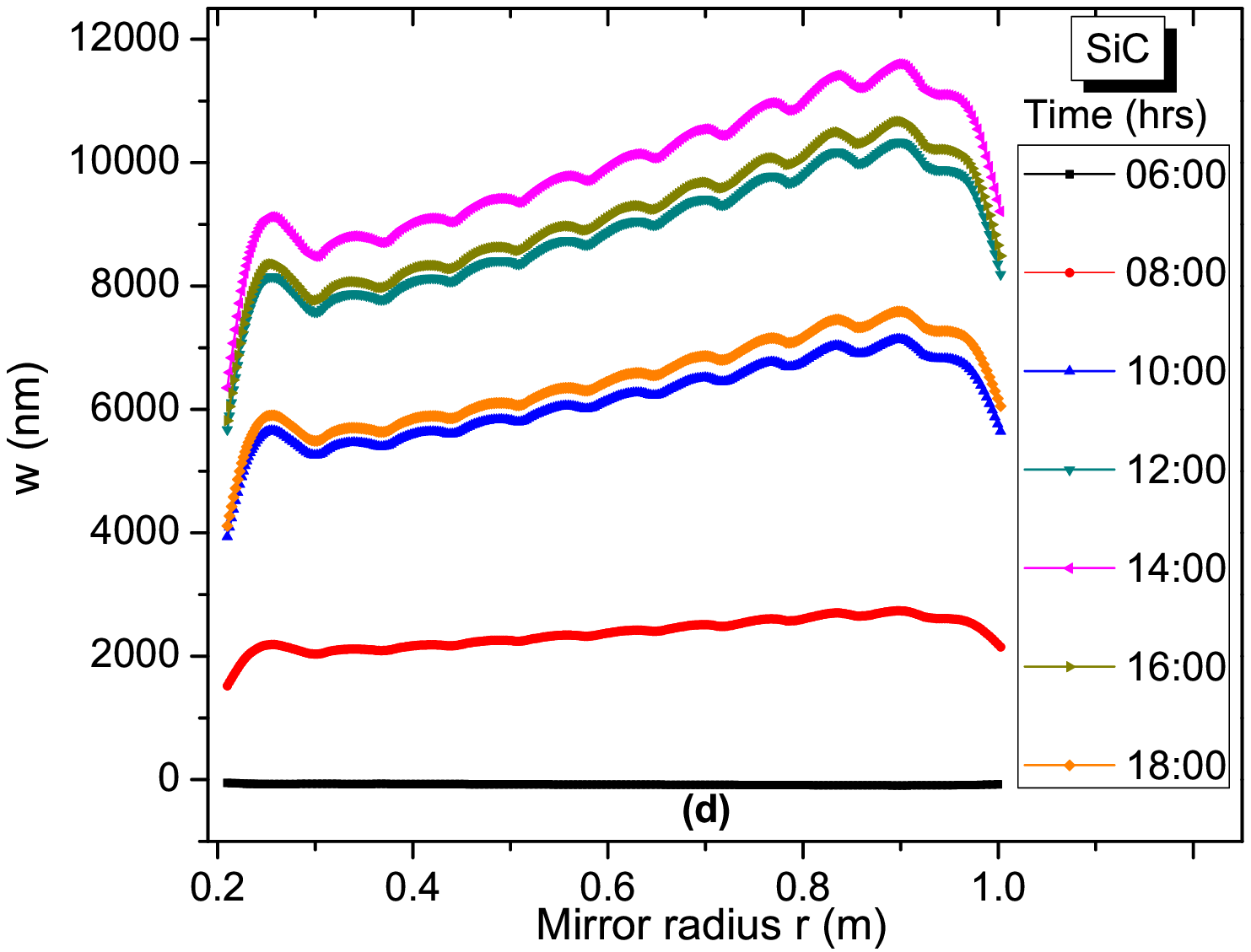}\\
\caption{ Thermally induced structural deformation: (a) for the Zerodur mirror and the corresponding z-displacements $w$ (nm) along the radial line $AA^\prime$ of the mirror faceplate for (a) Zerodur, (b) for Pyrex and (c) for SiC at different times of the day. The ambient temperature range was 20-30$^\circ$C and $h=5$~W/m$^2$.}
\end{figure}
 The maximal departure of $\Delta T_{{\scriptsize\textrm{max}}}$, $\delta T_{{\scriptsize\textrm{max}}}$   and $\xi T_{{\scriptsize\textrm{max}}}$ for an ambient temperature range $T_{\textrm{a}}:20^\circ$C$-30^\circ$C for Zerodur, Pyrex and SiC material is listed in Table 2. From the Table 2, we may note that the amplitude of both  $\Delta T$ and $\delta T$ depend crucially on the thermal conductivity of the material and the convective heat transfer coefficients used in simulations. A more efficient cooling mechanism is necessary, especially in Zerodur and Pyrex, to remove the excessive heat to minimize the mirror seeing effects.

 A diurnal variations of $\delta T$ computed between two reference points located at the rib wall and the center of hex-cell is shown in Fig.~4. Even though the surface temperature is always above the ambient, the impact of  $\Delta T$ and $\delta T$ are manifestly more pronounced during the noon hours.

\subsection{Thermally induced surface deformation}
 The non-uniform temperature field such as shown in Fig.~3(a), causes the mirror substrate to deviate away from its nominal defined shape. Thermally induced shape change can be highly irregular in lightweighted structures.  The surface distortions for each temperature input field were computed by solving the FEM structural model. The reference temperature $T_{\small\textrm{ref}}$ for the undeformed mirror geometry is assumed to be $20^\circ$C. The impact of lightweighted hex-cell geometry is clearly noticeable in the simulated image showing the surface deformation in Fig.~5(a) for Zerodur mirror. The corresponding axial displacements $w$ along the radial line $AA^\prime$ of the mirror faceplate for different substrate materials are also plotted in Figs.~5(b)--5(d). The surface displacement $w$ is positive in Pyrex and SiC substrates and gradually increases in radially outward direction.  However, for Zerodur material, $w$ can be both positive or negative because of zero crossing of the CTE around $25^\circ$C. For a given temperature field, the surface undulations, represented by $\delta w$,  are not uniform over the entire mirror surface but exhibit same periodicity as the underlying hex-cell structure. The displacement close to the mirror edges is expressed by $\xi w$.  The gradient across the mirror surface is approximated by the slope $w^{\prime}=dw/dr$.  The numerical values computed for  $\delta w$, $\xi w$ and $w^{\prime}$ during the peak mirror temperature at $t= 14:00$ are summarized in Table 3.

\begin{table}[htb]
 \centering\caption{Parameters describing the mechanical deformation of the surface.}\label{tab1}\small
 \begin{tabular}{|l|c|c|c|c|c|c|}
         \hline
           &  \multicolumn{3}{c|}{$T_{\textrm{a}}$ range:  $20^\circ$C$-30^\circ$C and $h$=5W/m$^2$} \\
           \cline{2-4}
            Material & \multicolumn{1}{c|}{$\delta w(\textrm{nm})$}   & \multicolumn{1}{c|}{$\xi w(\textrm{nm})$} & \multicolumn{1}{c|}{$ dw/dr\;(\times10^{-9})$} \\ \hline
         Zerodur      & 0.3-0.5     & 11.9    & -7.85   \\
         Pyrex        & 210-450      & 3773    & 3463.6  \\
         SiC          & 328-570     & 5248    & 5169.4    \\ \hline
        \end{tabular}
\end{table}

\begin{figure}[htb]
\centering\includegraphics[width=6.3cm]{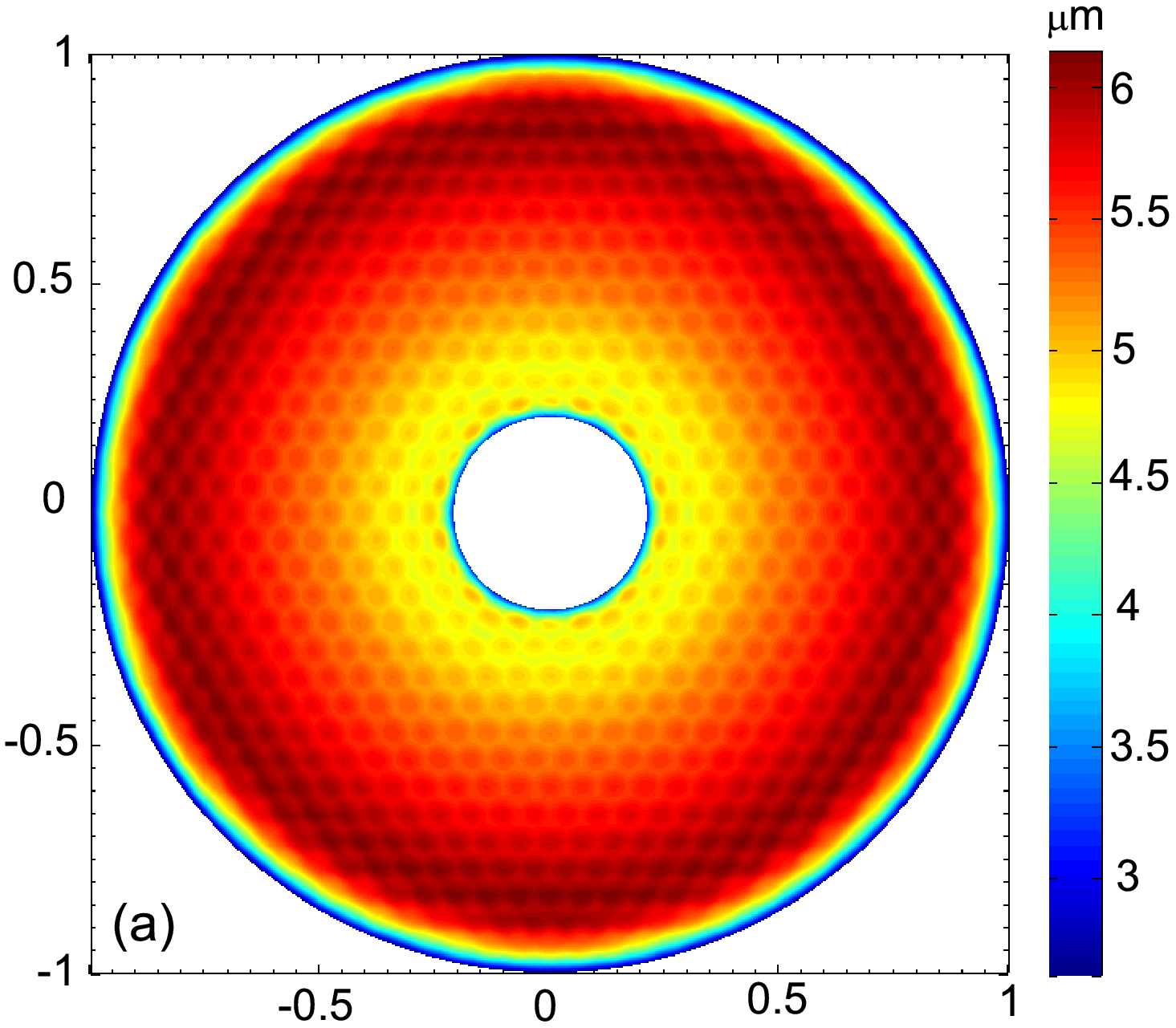}
\centering\includegraphics[width=6.6cm]{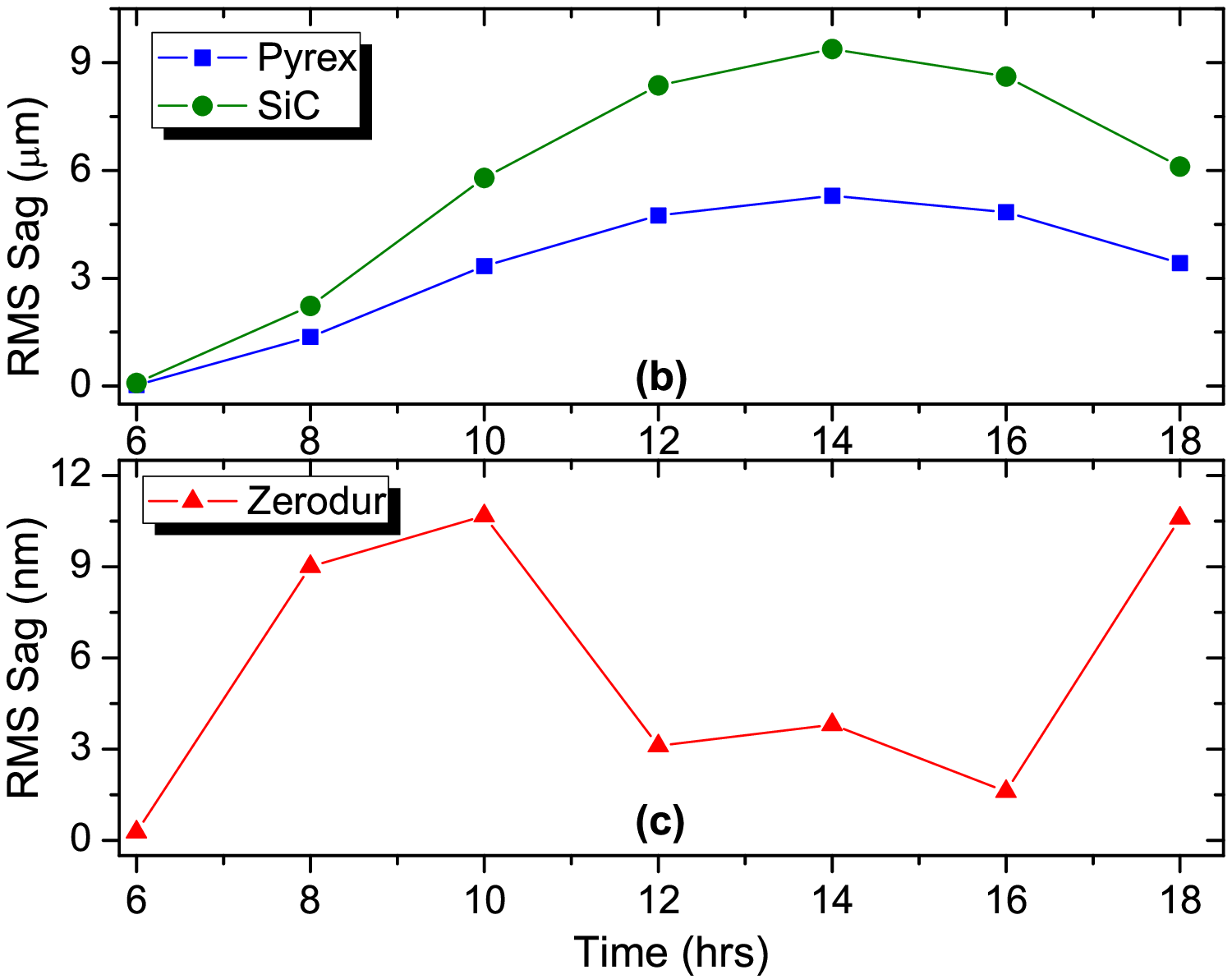}
\caption{(a) Surface sag for thermally deformed Pyrex mirror at $t=14$ hrs and $h=5$~W/m$^2$. The variation of RMS surface sag with daytime (b) for Pyrex and SiC and (c) for Zerodur mirror.}
\end{figure}
 The FEM studies were also performed at other temperature range of interest, i.e., $0^{\circ}$C$-10^{\circ}$C and $10^{\circ}$C$-20^{\circ}$C. The  surface deformation data was found almost similar to that given in Table 3.

\section{Optical analysis}
\subsection{Surface sag and Zernike polynomials}
 The thermally induced deformation data obtained from FEA in the previous section has to be transformed into a useful form to evaluate the optical performance of the mirror. A convenient way to model the deformation of an optical surface is to convert the displacement data to surface sag values. The  finite element computed node displacement of an arbitrary point on the mirror surface can be expressed as $d=(u^2+v^2+w^2)^{1/2}$, where $u$, $v$ and $w$ are components of displacement vector $d$ along $x,\;y$ and $z$ direction, respectively.  For small perturbations, the surface sag change $\Delta s$ and the node displacement at a radial point $r_{o}$ can be related by the following formula \cite{doy}:
\begin{equation}
\Delta s=w-\frac{\partial w(r_{0})}{\partial r}\sqrt{u^2+v^2}
\end{equation}

 The sag data for the axisymmetric mirror segment was imported into MATLAB and converted into full aperture using co-ordinate transformation of the form:\\
      \begin{equation}
      \left[\begin{array}{c}
        x^\prime \\
        y^\prime  \\
        z^\prime \\
        \Delta s^\prime
      \end{array}\right]
    =
       \left[ \begin{array}{cccc}
          \cos\theta & -\sin\theta & 0 & 0\\
          \sin\theta & \cos\theta & 0 & 0\\
          0 & 0 & 1 & 0\\
          0 & 0 & 0 & 1\\
        \end{array}
      \right]
     \left[\begin{array}{c}
        x \\
        y  \\
        z  \\
        \Delta s
      \end{array}\right]
      \end{equation}
        where $\theta=N\;\pi/3$ and $N=2,3,4,5$ and $6$. Equation(11) maps the surface sag data from the original mirror segment (unprimed coordinate) to other (primed coordinate) sectors.

The MATLAB function \texttt{TriScatteredInterp} which makes use of Delaunay triangulation algorithm, was used for interpolating irregularly spaced sag data over an uniformly sampled array of points. A typical example of surface sag for full aperture Pyrex mirror is shown in Fig.~6(a). The sag values vary between $2.8\mu$m$-6.13\mu$m and the root mean square (RMS) sag over the full aperture is $5.3\mu$m. The RMS wavefront error is twice the RMS sag. The diurnal variation of RMS sag for Pyrex, SiC and Zerodur, shown in Figs.~6(b) and 6(c), is consistent with variation of their CTE under specified temperature range.
\begin{figure}[htb]
\centering\includegraphics[width=6.5cm]{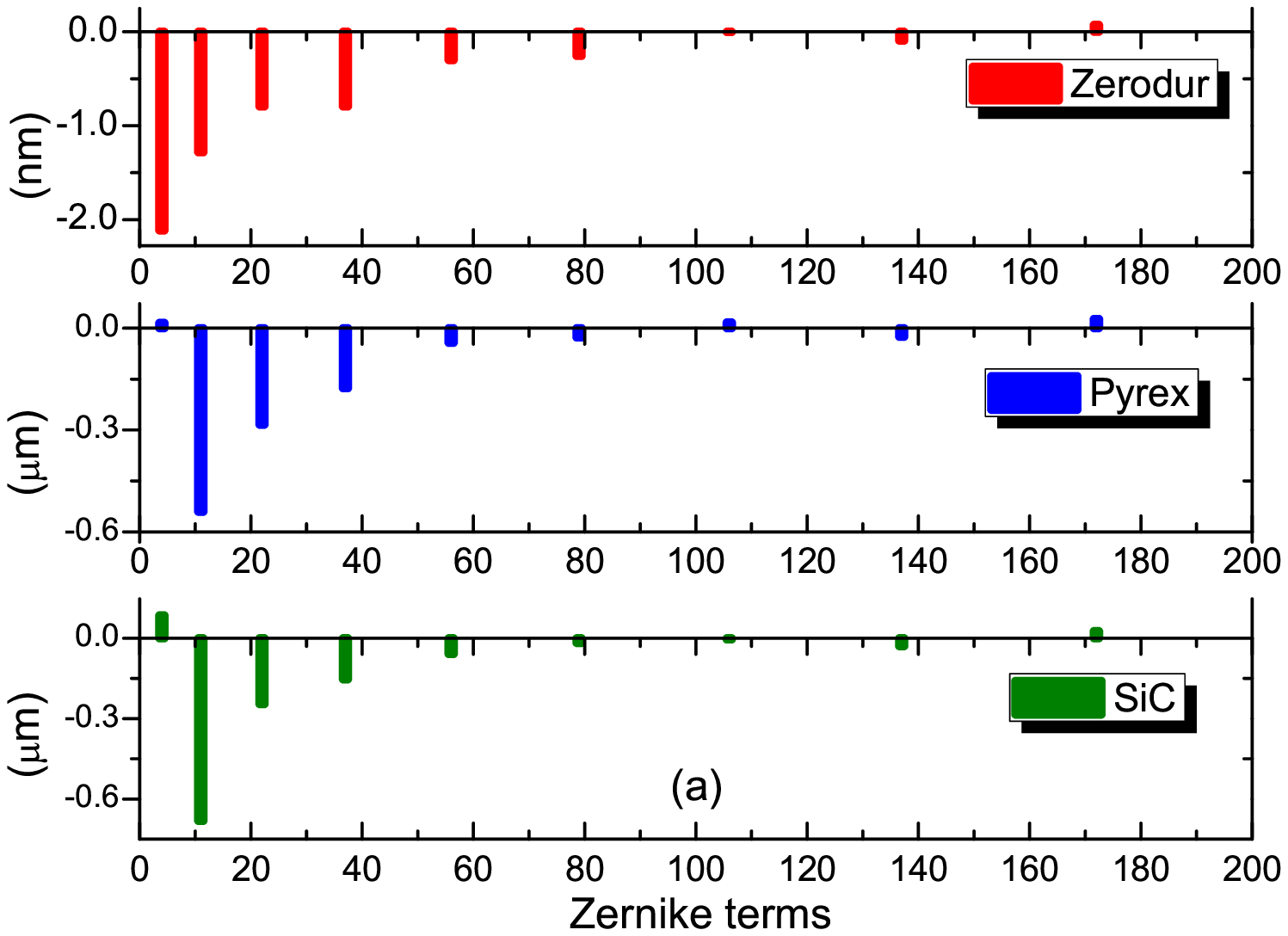}
\centering\includegraphics[width=6.7cm]{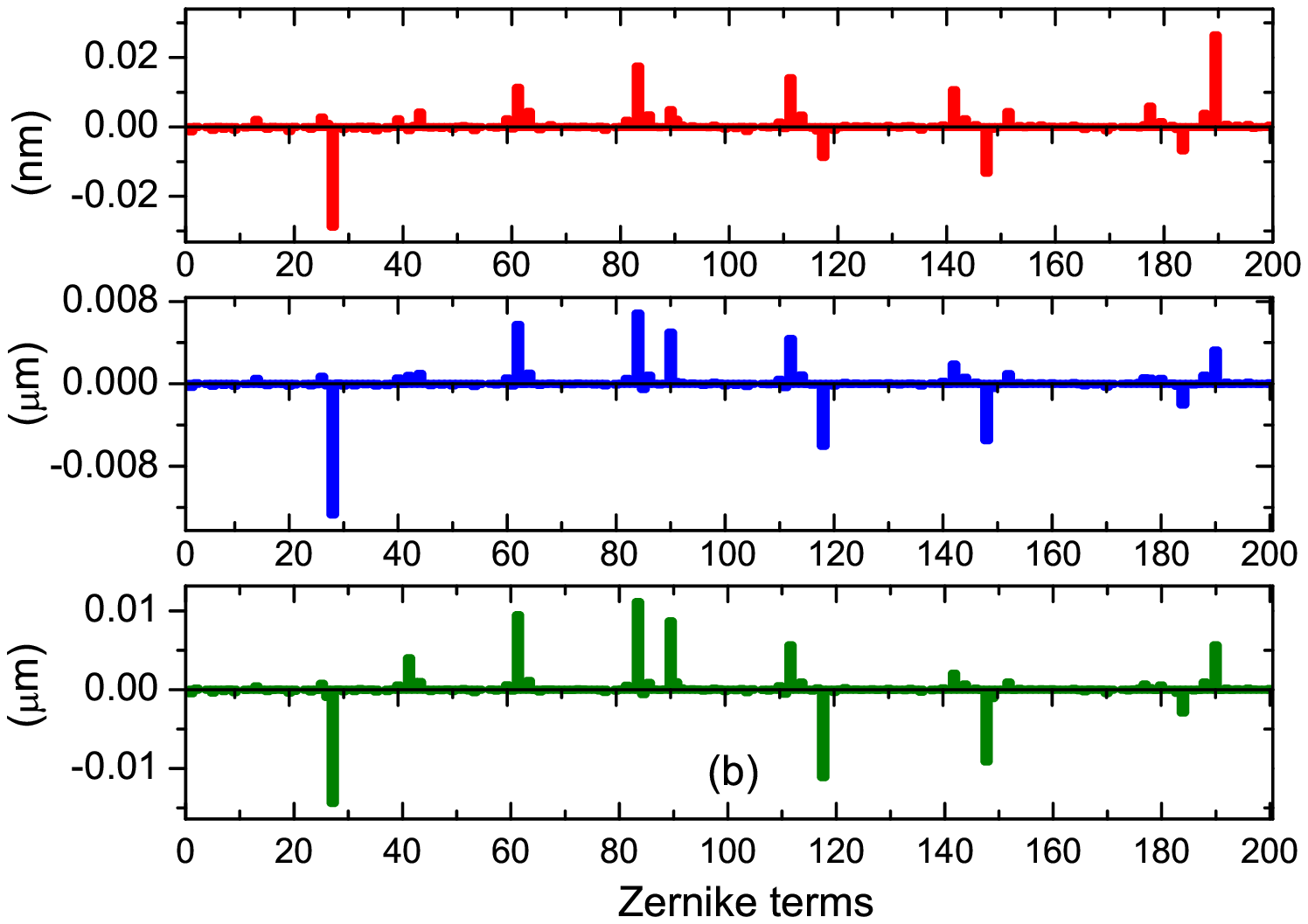}
\caption{Zernike expansion coefficients corresponding to (a) defocus, primary, secondary and other higher order spherical aberrations which contribute to overall all figure change and (b) the remaining low-amplitude terms which contribute to mid-frequency surface errors.}
\end{figure}
 The optical aberrations of the thermally deformed surface were further analyzed by fitting Zernike annular polynomials to 2D sag data. The main advantages of Zernike polynomials arise due to their orthogonality over the unit circle, rotational invariance, completeness property and direct relationship of expansion coefficients with known aberrations of the optical system \cite{nol,mah}. Each Zernike term is a product of three components, namely, a normalization factor, a radial part and an azimuthal part of type $\cos m\phi$ or $\sin m\phi$. The surface sag $\Delta s(\rho,\phi;\epsilon)$ for an annular pupil with obscuration ratio $\epsilon$, radial variable $\rho$  and azimuth angle $\phi$  can be written in terms of Zernike annular polynomial as:
\begin{equation}\label{eqs}
  \Delta s(\rho,\phi;\epsilon)=\sum_{j=1}^{\infty}\;a_{j}\;Z_{j}(\rho,\phi;\epsilon)
\end{equation}
where $j$ is a single index polynomial-ordering term  and $a_{j}$ are expansion coefficients \cite{my}.
\begin{figure}[htb]
\centering\includegraphics[width=6.4cm]{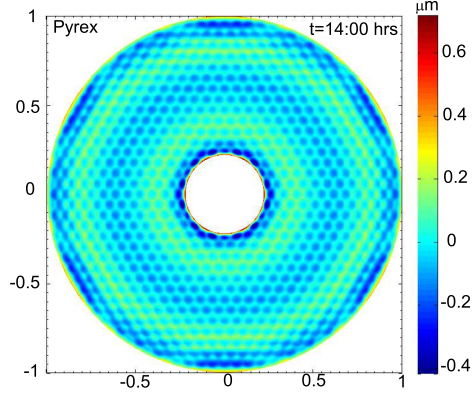}
\caption{Residual map obtained by subtracting the reconstructed Zernike surface from the original sag data.}
\end{figure}

The $j$-th order Zernike polynomials $Z_{j}(\rho,\phi;\epsilon)$ in Eq.~(12) is defines as \cite{mah}
\begin{equation}
  Z_{j}(\rho,\phi;\epsilon) = \left\{
\begin{array}{l l l}
  \sqrt{2(n+1)}\;R_n^m(\rho;\epsilon) \cos m\phi & \quad \mbox{$m\neq0$,  $j$ even}\\
  \sqrt{2(n+1)}\;R_n^m(\rho;\epsilon) \sin m\phi & \quad \mbox{$m\neq0$, $j$ odd}\\
    \sqrt{n+1}\;R_n^0(\rho;\epsilon) & \quad \mbox{$m=0$} \end{array} \right.
\end{equation}
where $n$ is  radial degree and $m$ is azimuthal frequency such that $ m\leq n$ and $n-|m|$ is even. The procedure to obtain expressions for radial component $R_n^m(\rho;\epsilon)$ in Eq.~(13) is discussed in Ref \cite{mah}.
\begin{table}[htb]
\centering\caption{Low-order radial Zernike aberration terms.}\label{tab5}\small
\begin{tabular}{r|l|l|l}
  \hline
$j$  & Radial terms & Expression & Meaning \\
  \hline
  1  & $R^0_0(\rho;\epsilon)$ & 1 & Piston \\
  4  & $R_2^0(\rho;\epsilon)$ & $\frac{\sqrt{3}(\epsilon^2-2\rho^2+1)}{\epsilon^2-1}$  & Defocus \\
  11 & $R_4^0(\rho;\epsilon)$ & $\frac{\sqrt{5}(6\rho^4-6\rho^2+\epsilon^4+\epsilon^2(4-6\rho^2)+1)}{(\epsilon^2-1)^2}$ & Pri. spherical \\
  22 & $R_6^0(\rho;\epsilon)$ & $\frac{\sqrt{7}(-20\rho^6+30\rho^4-12\rho^2-3\epsilon^4(4\rho^2-3)+3\epsilon^2(10\rho^4-12\rho^2+3)+\epsilon^6+1)}{(\epsilon^2-1)^3}$ & Sec. spherical  \\
    \hline
\end{tabular}
\end{table}

We used MATLAB based open-source toolkit developed at the University of Arizona, for the Zernike analysis of the sag data \cite{kim}. The computed Zernike coefficients for three substrate materials are plotted in a bar chart shown in Fig.~7. Based on their relative amplitude,  various Zernike terms are separated into two parts. The leading contribution comes from purely radial part (i.e., $m=0$ case in Eq.~(13)) plotted in Fig.~7(a). These terms correspond to piston ($j=1$), defocus ($j=4$), primary spherical ($j=11$), secondary spherical ($j=22$,) and higher order ($j=37,56,79,106,137,172$) spherical aberrations that are responsible for driving the overall figure change of the mirror. Among these, the defocus term (Fig.~7) dominates for Zerodur, while primary spherical  aberration dominates for Pyrex and SiC.  The first 4 radial polynomials and their associated meanings are given in Table 4. The amplitude of these terms diminishes progressively with increasing order $j$. The pure radial and few low-frequency azimuthal terms are also reported to have strong influence on ocular aberrations encountered in vision related problems \cite{app}.
\begin{figure}[htb]
\centering\includegraphics[width=7cm]{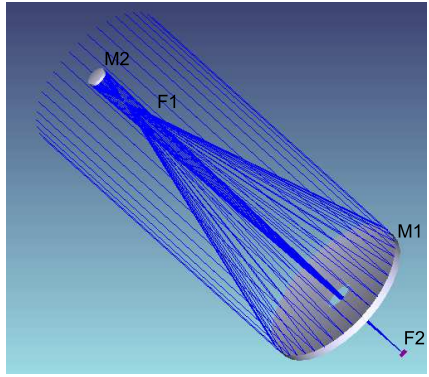}
\caption{Optical model of a Gregorian-type reflecting telescope in Zemax. M1: primary mirror; M2: secondary mirror; F1: primary focus; F2: secondary focus or image plane.}
\end{figure}

\begin{table}[htb]
\centering\caption{Predominant low-amplitude Zernike terms consisting of azimuth and radial parts.}\label{tab5}\small
\begin{tabular}{r|l}
  \hline
$j$  & Expressions  \\
   \hline
  \rule{0pt}{3ex}  28  & $\sqrt{14}\,R^6_6(\rho;\epsilon)\cos6\phi$  \\
  \rule{0pt}{3ex} 62  & $\sqrt{22}\,R_{10}^6(\rho;\epsilon)\cos6\phi$  \\
  \rule{0pt}{3ex}84  & $\sqrt{26}\,R_{12}^6(\rho;\epsilon)\cos6\phi$ \\
  \rule{0pt}{3ex}112 & $\sqrt{30}\,R_{14}^6(\rho;\epsilon)\cos6\phi$   \\
  \rule{0pt}{3ex}118 & $\sqrt{30}\,R^{12}_{14}(\rho;\epsilon)\cos12\phi$ \\
  \rule{0pt}{3ex}148 & $\sqrt{34}\,R_{16}^{12}(\rho;\epsilon)\cos12\phi$  \\
  \rule{0pt}{3ex}184 & $\sqrt{38}\,R_{18}^{12}(\rho;\epsilon)\cos12\phi$ \\
  \rule{0pt}{3ex}190 & $\sqrt{38}\,R_{18}^{18}(\rho;\epsilon)\cos18\phi$  \\
  \hline
\end{tabular}
\end{table}

The remaining low-amplitude Zernike terms are plotted in the Fig.~7(b). The resulting amplitude of these mixed  (product of radial and angular) Zernike terms is about 2-order smaller than the pure radial terms. The predominant low-amplitude terms comprising radial and angular part are given in Table 5. Ordinarily the higher order terms are not identified with any specific optical aberration. The solitary presence of $\cos6\phi$ and its harmonics can be associated with the shape of thermal print-through errors arising from the 6-fold symmetry of the hexagonal pockets. It is to be noted that amplitude of these terms (see Fig.7(b)), does not decline appreciably with increasing $j$. This implies that using more number of Zernike terms is unlikely to improve the fitting accuracy of the sag data further. One reason for such behavior is the limited ability of the Zernike polynomials to capture high-frequency surface imperfections  as oppose to their effectiveness in capturing low-frequency global changes in the surface shape. That is why in Zernike representation, the impact of high frequency thermal print-through on surface sag is likely to be swamped by low-order figure changes induced by defocus and all spherical terms. In order to effectively visualize small-scale surface elevations, we computed residual sag by extracting the Zernike reconstructed surface from the original sag data. The mid-spatial frequency features, typical of thermal-print through affects, are clearly seen in the residual map shown in Fig.~8.
\begin{figure}[htb]
\centering\includegraphics[width=9.5cm]{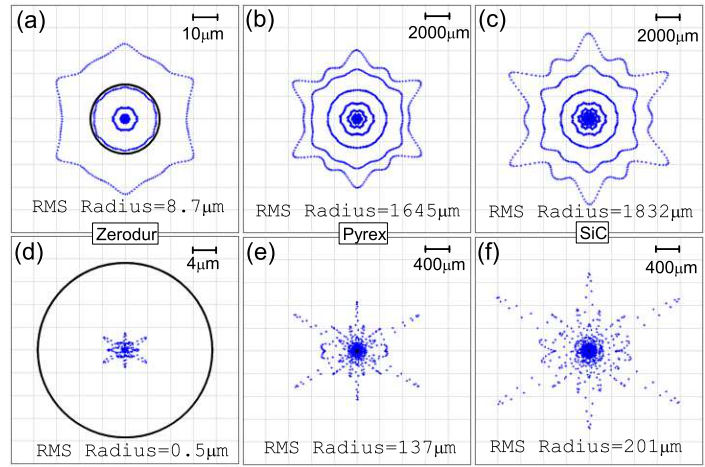}

\caption{The spot diagram showing the telescope performance for a on-axis $0^\circ$ field point. The blue traces represent the ray distribution in the image plane due to temperature induced surface imperfections in M1. The pattern has an envelop of six-fold symmetry due to hexagonal nature of the elements, as expected. The diffraction limited Airy disc is indicated by the black circle. The specified wavelength is 0.55 $\mu \textrm{m}$ and the radius of the airy disc is close to 15 $\mu \textrm{m}$. Top row (a)--(c), when all the Zernike terms were included. Bottom row (d)--(f), without piston, focus and spherical terms.}
\end{figure}
\subsection{Image quality}
 The surface aberration data of thermally deformed mirror was analyzed using Zernike polynomials and the optical performance was evaluated in a commercial ray tracing software Zemax.  A simple telescope design was created in Zemax as shown in Fig.~9. The design conforms to optical prescription of M1 given in Section 3. The aberration data can be incorporated into the optical model of the telescope by assigning a suitable surface type --the \emph{Zernike Standard Sag}, to the M1 in Zemax. This surface type is commonly used for describing mechanical deformations. Selecting \emph{Zernike Standard Sag} surface allows user to define the standard surface (e.g. plane, spheres, conics etc) plus additional Zernike terms that can be imposed on the standard surface to represent surface aberrations \cite{zem}. The aberration data for M1 along with number of Zernike terms, normalized radius of the aperture and a list of externally computed Zernike coefficients was directly imported into Zemax using  Extra Data Editor. The M2 was assumed to be free from any surface irregularity.

 In Zemax, there are many ways to assess the optical performance of the system. We used spot diagrams and modulation transfer function to evaluate the performance of the telescope system. The thermal distortions are assumed to become worse when the mirror temperature reaches its maximum at $t=14:00$ hrs. For the worst case scenario, the effect of thermally induced aberrations on the telescope image quality can be readily seen in the spot diagram shown in Fig.~10. Except for Zerodur material, the RMS spot radius (Figs.~10(a)-10(c)) for Pyrex and SiC is significantly larger than the size of the airy disc, implying the diffraction limited performance of M1 made of high CTE materials, is severely compromised by thermal heating. The contribution of low amplitude Zernike terms (Fig.~7(b)) is estimated by setting the radial terms (Fig.~7(a)) to zero in Zemax. Clearly, the low amplitude terms add significant scatter to the energy distribution as shown in Figs.~10(d)-10(f). The envelop of the spot scatter is also indicative of the periodicity and shape of the `surface bumps' caused by uneven temperature distribution resulting from underlying hex-cell structure of the M1. Another way to examine the optical performance is to compute the contrast variations of the imaging system at different spatial frequencies. The variation of contrast, i.e., MTF for Pyrex and SiC is plotted in Figs.~11(a) and 11(b), respectively. Considering the severity of the thermal distortions, a steep fall in contrast at relatively low-frequencies is not completely unexpected.

 The preceding studies were also repeated for convective heat transfer coefficient $h=15$~W/m$^2$. The corresponding RMS spot radii for the Zerodur, Pyrex and SiC were found to be $3.3\mu$m, $1305\mu$m and $1798\mu$m. This is only a moderate improvement. This shows that forced circulation of  ambient air alone in high CTE materials (Pyrex and SiC) is not enough to keep the thermal distortions within tolerable limits. For effective thermal control, the temperature of the circulated air has to regulated few degrees below the ambient temperature. A detailed CFD analysis is however needed to arrive at the optimal coolant air temperature that would minimize the mirror seeing and thermal distortions of the M1. The numerical studies presented in this paper would complement those efforts. Finally, we draw the attention of the scientific community to consider the present scheme of opto-thermal analysis for optimization of the solar telescope design.

\begin{figure}[htb]
\centering\includegraphics[width=9.5cm]{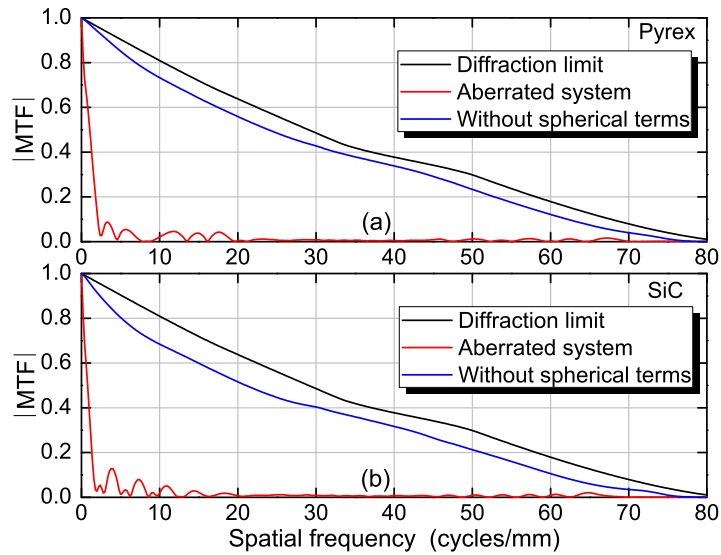}
\caption{The system contrast in terms of transverse modulation transfer function at various spatial frequencies for a $0^\circ$ field angle (a) for Pyrex and (b) for the SiC mirror. For Zerodur (not shown), the ray aberrations are well within the diffraction limits and three MTF curves would overlap.}
\end{figure}

\section{Conclusions}
The temperature stabilization of a primary mirror system is one of the most challenging tasks in building a large aperture solar telescope.  To achieve the desired optical performance of the telescope, the thermal response of the primary mirror needs to be accurately predicted and suitably controlled under the harsh observing conditions. For large aperture telescope, the lightweighted mirror offer significant advantages over traditionally used solid blanks. However, the pocketed cell structure created by the lightweighting process leads to geometry dependent material inhomogeneities and non-uniform temperature patterns. Besides providing physical support to the faceplate, the rib walls of the lightweighted structure also serve as discrete channels for the heat flow, giving rise to periodic thermal print-through errors across the mirror surface. We carried of a detailed thermo-optic analysis of a 2m class solar telescope mirror. A time dependent heat transfer and structural model of the lightweighted mirror was solved using FEA methods. The mirror faceplate was heated by a radiant energy from the sun. The distinct appearance of thermal print-through patterns markedly depends on the thickness of the faceplate and changing thermal environment of the surroundings. The thermal displacement FEA data  was converted to surface sag values useful for evaluating the optical quality of the telescope. The Zernike analysis of the sag data showed that while the major contribution to the optical figure error comes from primary, secondary and higher order spherical aberration terms, the thermal print-through arising from underlying hex-cell geometry mainly contributes to low amplitude, mid-spatial frequency surface errors. A high scatter in ray-traced RMS spot size and a steep fall in MTF signifies a severe degradation in optical image quality of high CTE (SiC and Pyrex) materials.
While SiC and Pyrex materials show a greater propensity for thermal distortions. But that alone may not be a sufficient ground to completely rule out their usages at least in moderate sized mirrors. It is also true that SiC and pyrex may not be able to match the performance level of ULE materials, but in principle, the structural instability can be better ensured by designing an efficient temperature control system.  A higher convective heat transfer coefficient showed only a marginal improvement in the image quality which also reinforces the need for a better cooling mechanism to remove the excessive heat from the M1. The approach outlined in this paper would be useful to design an effective thermal control and temperature stabilization system for solar observations.
\section*{Acknowledgment} We would like to thank Dr. Ping Zhou at Large Optics Fabrication and Testing Group, Arizona University, for suggesting to use a faster and more versatile MATLAB code `SAGUARO' for the Zernike analysis.


\begin{thebibliography}{99}
\bibitem{my} We have chsoen the normalization factor  and  the indexing scheme similar to Noll's notaion which is also followed in Zemax ray-tracing software.
\end{thebibliography}
\end{document}